\RequirePackage{fix-cm}

\documentclass[smallextended]{svjour3}       
\smartqed  
\usepackage{longtable}
\usepackage[english]{babel}
\usepackage[T1]{fontenc}
\usepackage[utf8]{inputenc}
\usepackage{booktabs} 
\usepackage{graphicx}
\usepackage{scrtime}
\usepackage{comment}
\usepackage[dont-mess-around]{fnpct}
\usepackage{authblk}
\usepackage{etoolbox}
\newcommand*{\affaddr}[1]{#1} 
\begin{document}

\title{Ethics of AI: A systematic literature review of principles and challenges}

\titlerunning{Ethics of AI: A SLR of principles and challenges}        

\author{Arif Ali Khan \and Sher Badshah \and Peng Liang \and Bilal Khan \and   Muhammad Waseem \and Mahmood Niazi \and Muhammad Azeem Akbar}

\authorrunning{Khan et al.} 

\maketitle

\begin{abstract}
Ethics in AI becomes a global topic of interest for both policymakers and academic researchers. In the last few years, various research organizations, lawyers, think tankers and regulatory bodies get involved in developing AI ethics guidelines and principles. However, there is still debate about the implications of these principles. We conducted a systematic literature review (SLR) study to investigate the agreement on the significance of AI principles and identify the challenging factors that could negatively impact the adoption of AI ethics principles. The results reveal that the global convergence set consists of 22 ethical principles and 15 challenges. Transparency, privacy, accountability and fairness are identified as the most common AI ethics principles. Similarly, lack of ethical knowledge and vague principles are reported as the significant challenges for considering ethics in AI. The findings of this study are the preliminary inputs for proposing a maturity model that assess the ethical capabilities of AI systems and provide best practices for further improvements. 
\keywords{AI ethics, \and   machine ethics \and principles \and challenges \and systematic literature review}
\end{abstract}

\section{Introduction}
\label{sec:intro}

Artificial intelligence (AI) technologies are considered important across a vast array of industries including health, manufacturing, banking and retail [1]. However, the promises of AI systems like improving productivity, reducing costs, and safety has now been considered with worries, that these complex systems might bring more ethical harm than economical good [1].

Artificial intelligence (AI) and autonomous systems have a significant effect on the development of humanity [2]. The autonomous decision-making nature of these systems raises fundamental questions i.e., what are the potential risks involved in those systems, how these systems should perform, how to control such systems and what to do with AI-based systems? [2]. Autonomous systems are beyond the concepts of automation by characterising them with decision-making capabilities. The development of autonomous system components, such as intelligent awareness and self-decision making is based on AI concepts.

There is a political and ethical discussion to develop policies for different technologies including nuclear power, manufacturing etc. to control the ethical damage they could bring. The same ethical potential harm also exits in AI systems and more specifically they might end human control [2].The real-world failure and misuse incidents of AI systems bring the demand and discussion for AI ethics [3]. The ethical studies of AI technologies revealed that AI and autonomous systems should not be only considered a technological effort. There is a broad discussion that the design and use of AI-based systems are culturally and ethically embedded [4]. Developing AI-based systems not only need technical efforts but also include economic, political, societal, intellectual and legal aspects [4]. These systems significantly impact the cultural norms and values of the people [4].
The AI industry and specifically the practitioners should have a deep understanding of ethics in this domain. Recently, AI ethics get press coverage and public voices, which supports significant related research [3]. However, the topic is still not sufficiently investigated both academically and in the real-world environment [4]. There are very few academic studies conducted on this topic, but it is still largely unknown to AI practitioners. The Ethically Aligned Design (EAD) [6] guidelines of IEEE mentioned that ethics in AI is still far from being mature in an industrial setting [5]. The limited or no knowledge of ethics for the AI industry develop the gap, which indicates the need for further academic and in practice research.  

The aim of this study is to conduct a systematic literature review (SLR) and explore the available literature to identify the AI ethics principles. Moreover, the SLR study uncover the key challenging factors that are the demotivators for considering the ethics of AI. The following research questions are developed to achieve the given core objectives:
    \begin{itemize}
        \item RQ1: What are the key principles of AI ethics?
        \item RQ2: What are the challenges of adopting ethics in AI?
    \end{itemize}
    
The remaining content of paper is structured as follow: Section \ref{sec:Background} presents the background of the study and the research methodology is reported in Section \ref{sec:Methodology}. The SLR data are provided in Section \ref{sec:Reporting the review} and results and analysis are discussed in Section \ref{sec:Details results and analysis}. Finally, Section \ref{sec:Threats to validity} provides an overview of threats to the validity of the study and Section \ref{sec:conclusions} conclude the findings with future directions.

\section{Background}
\label{sec:Background}
The implementation of AI or machine intelligence concepts brings a technological revolution that change both science and society. Human to machine power transformation sparked important societal debate about the principles and policies that guide the use and deployment of AI systems [7]. Various organizations have developed ad hoc committees to draft the policy documents for AI ethics. These organizations reportedly developed AI policies and guide documents [7]. In 2018, technology corporates such as SAP and Google publicly introduced guidelines and policies for AI-based systems [7]. Similarly, Amnesty International, the Association of Computing Machinery (ACM) and Access Now comes up with principles and recommendations for AI technologies. The Trustworthy AI European Commission’s guidelines were developed with the aim to promote lawful, ethically sound and robust AI systems [8]. The report “Preparing for the Future of Artificial Intelligence” prepared by the Obama administration’s presents a thorough survey that focuses on the current AI research, its applications and impact on society [9]. The report further presents recommendations for future AI related actions. The “Beijing AI Principles” guidelines [10] proposed various principles in the domain of AI research, development, use and governance. These principles present a framework that focus on AI ethics. 

The world largest technical professional organization, IEEE launches the guidelines Ethically Aligned Design (EAD) [6] that provides a framework to address the ethical and technical values of AI systems based on a set of principles and recommendations. The EAD framework consists of the following eight general principles to guide the development and implementations of AI-based systems: human rights, well-being, data agency, effectiveness, transparency, accountability and awareness of misuse. Organizations such as ISO and IEC also embark on developing standard for AI [11]. ISO/IEC JTC 1/SC 42 is a joint ISO/IEC international standard committee that focus on the entire AI ecosystem including ethical and social concerns, standardization, AI governance, AI computational approach and trustworthiness [11]. The effort of different organizations to shape AI ethics not only determine the need of guidelines, tools, techniques, but also the interest of these organizations to manage ethics in a way that meet their respective priorities.

However, recently published studies reported that the existing guidelines developed for ethics of AI are not effective and adopted in practice [12]. It is evident from the empirical study conducted by McNamara et al. [13] to test the influence of the ACM code of ethics in the decision-making process of software development. The results of the study revealed that the ACM code of ethics have no impact in making ethical decisions. The lack of effective techniques makes it challenging to successfully scale the available guidelines into practice [12].	Vakkuri et al. [12] used the accountability, responsibility, and transparency (ART) framework [14] and developed the conceptual model to explore ethical consideration in the AI environment. The conceptual model is empirically validated by conducting multiple case studies. The empirical results are concluded by highlighting that AI ethics principles are still not in practice; however, some common concepts are considered such as documentation. Moreover, the study findings revealed that practitioners consider the social impact of AI systems [12].

There are no tools, methods or frameworks that fill the gap between the AI principles and their implementation in practice. Further studies in this area should conduct that explicitly discuss the AI ethics principles, challenges and provide evaluation standards/models that  guide AI industry to consider ethics in practice.   
\section{Research Method }
\label{sec:Methodology}
\begin{figure*}[!htbp]
  \centering
  \includegraphics[scale=0.71]{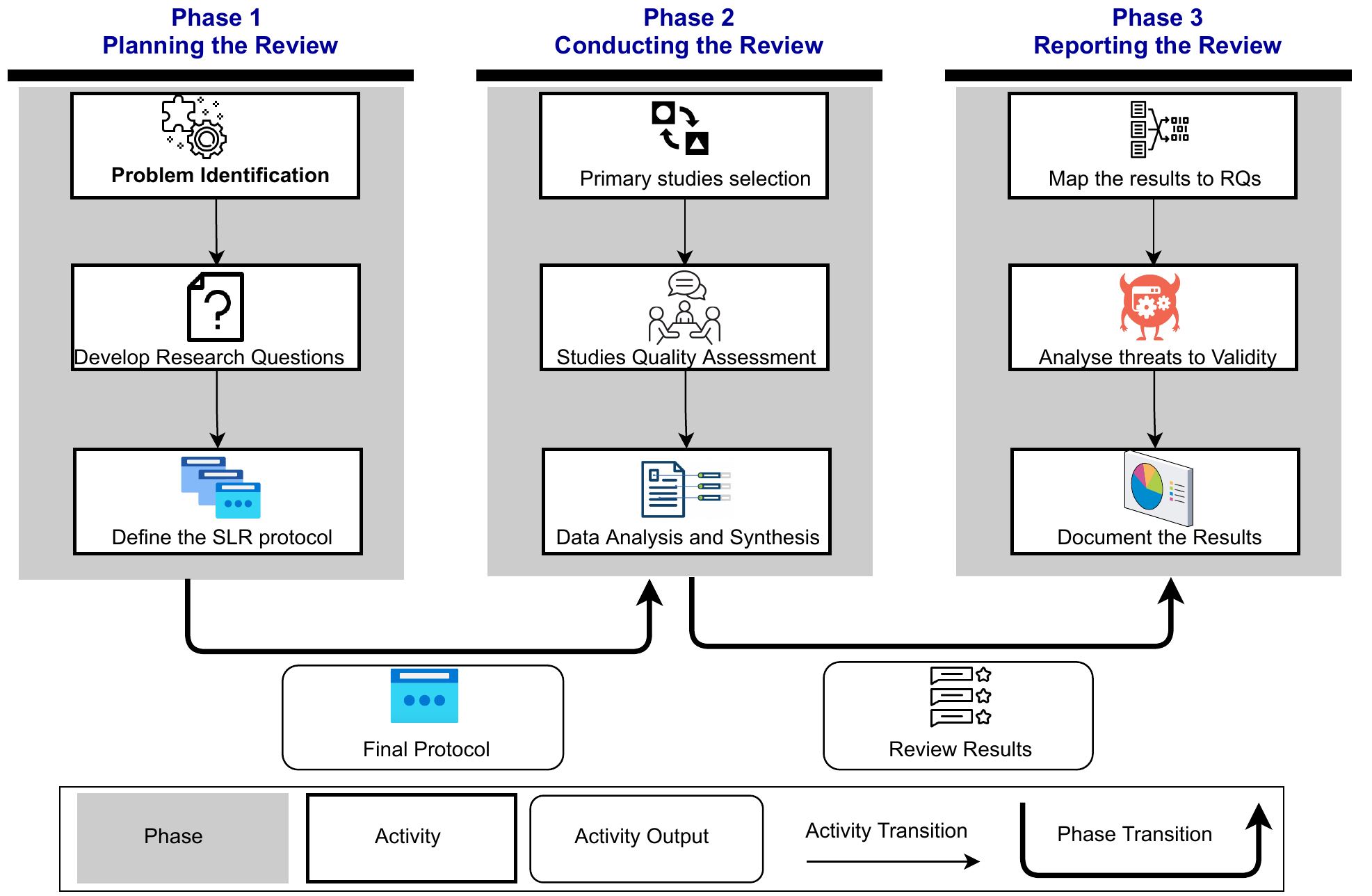}
  \caption{SLR research process}
  \label{fig:SLRproces}
\end{figure*}

Systematic literature review (SLR) approach is used to explore the available primary studies. SLR is a widely adopted literature survey method in evidence-based software engineering domain. SLR is “a means of evaluating and interpreting all available research relevant to a particular research question, topic area, or phenomenon of interest” [15]. The Kitchenham and Charters [15] SLR guidelines are used to conduct this study and systematically address the research questions. The SLR process plan is provided in Fig. \ref{fig:SLRproces} and thoroughly discussed in the following sections.

\subsection{Research questions (RQs)}
\label{sec:RQs}
Research questions development in SLR studies is the most significant phase [15]. Developing research questions require deep understanding of the research area in general and the research problem in specific. We primarily studied relevant articles [3-9] to better understand the problem and develop the questions of interest. The questions are finally developed based on the research concepts discussed in the mentioned research sources [3-9]. The details of the research questions are provided in Section \ref{sec:intro}. 

\subsection{Data sources}
\label{sec:Data Sources}
The authors had a series of team discussions to identify the list of digital data sources. The selected digital repositories are explored to extract the relevant data in order to address the given research questions (see Section \ref{sec:intro}). Finally, the following digital libraries are selected based on the authors SLR experience, discussions and guidelines provided by Chen et al. [16]: Springer Link, Science Direct, IEEE Xplore, Wiley Online Library and ACM Digital Library. These are the world leading digital data sources which collect a large number of original information and communication technology studies [16].

\subsection{Search strategy }
\label{sec:Search strategy}
The research questions are analysed by the second and third authors to extract the terms or keywords used for the search process. All the authors participated in the group discussion to finalise the search terms and retrieve the relevant data from the selected repositories. Pilot search terms and strings are made that finally contributed to develop the following agreed search string:
\begin{center} ("artificial intelligence ethics" OR "AI ethics" OR "machine learning ethics" OR "software ethics") AND (“resistance” OR “barriers” OR “limitations” OR “challenges”)\end{center}

The “principles” and “guidelines” terms were excluded from the final search string because these terms return irrelevant data from different other domains. The given search string was specifically testified during the pilot attempts to explore the data related to the AI “principles” and “guidelines” and we noticed that it precisely returns the desire results related to the RQ1, i.e., “principles” and “guidelines”.

The search terms are concatenated using “AND” and “OR” operators to develop the search strings. The selected digital repositories have a customised search mechanism. The search strings are executed using the personalised search mechanism of electronic data sources.  

\subsection{Inclusion/Exclusion criteria}
\label{sec:Inclu/Exclu}
The inclusion/exclusion criteria are developed to filter the search string findings and remove irrelevant, not accessible, redundant and low-quality studies. The criteria are developed by the first and fifth authors, which are finalised by all the authors in the regular consensus meeting (see Table \ref{tab:InclusionExclusion}).

\begin{table*}
  \caption{Inclusion/Exclusion criteria}
  \label{tab:InclusionExclusion}
  \begin{tabular}{cp{12.5cm}}
   \hline
    \textbf{No.} & \textbf{Inclusion criteria}\\
    \midrule
    In1 & Consider articles that specifically focus on AI ethics. \\
    In2 & Primary studies published in conferences, research workshops, book chapters, journals and magazines. \\
    In3 & Peer-reviewed and available in full text \\
    In4 & Written in the English language \\
    \midrule
     \textbf{No.} & \textbf{Exclusion criteria}\\
     \midrule
     Ex1 & If two studies are published from the same project, then exclude the one with minimum contribution. \\
     Ex2 & Exclude grey literature material. \\
     Ex3 & Remove duplicate studies. \\
     Ex4 & Discuss ethics in other domains. \\
     \bottomrule
  \end{tabular}
\end{table*}

\subsection{Study selection}
\label{sec:Study selection}
\begin{figure*}[!htbp]
  \centering
  \includegraphics[width=\linewidth]{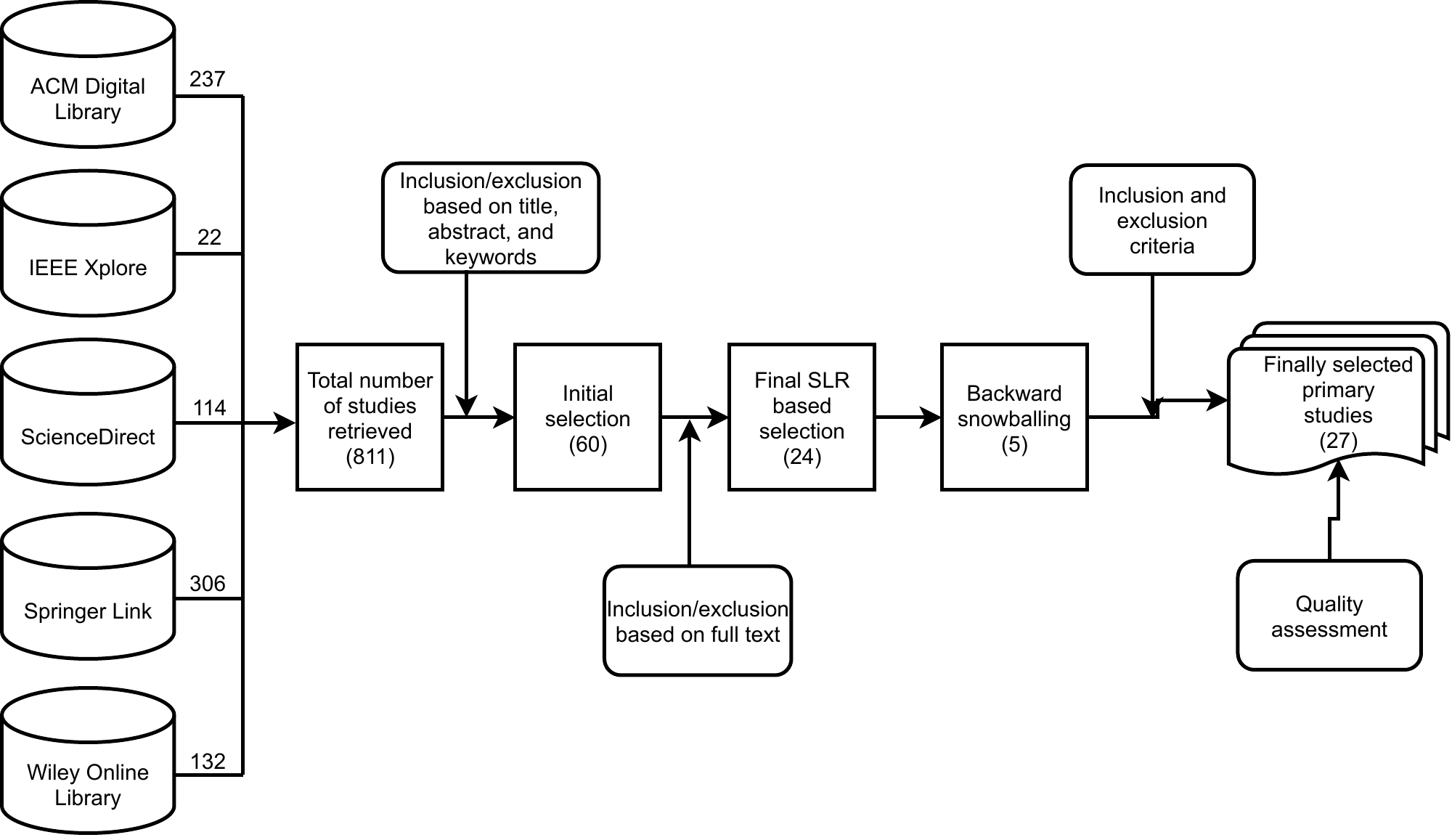}
  \caption{Studies selection process}
  \label{fig:StudyselectionProcess}
\end{figure*}

The search string discussed in Section \ref{sec:Search strategy} is used to explore the selected digital repositories. The search process was initiated on 23rd December 2020 and ended on 5th February 2021. The search string retrieved total 811 studies in the first phase, which were further filtered based on the study title, abstract and keywords (see Fig. \ref{fig:StudyselectionProcess}). In the second phase of the selection process, the inclusion/exclusion of the 60 studies are performed based on the full-text review. Finally, 24 primary studies are shortlisted using the SLR approach. Moreover, backward snowballing [17] is performed to search the references of the selected 24 studies. The backward snowballing is previously used by Tingting et al. [18] to explore text analysis techniques in software architectural and we used with the aim to explore the references list of the selected primary studies to identify relevant studies that are missed during the SLR process. Additionally, 5 studies are selected, which are further filtered using the inclusion/exclusion criteria (see Fig. \ref{fig:StudyselectionProcess}). Eventually, only 3 studies fulfil the selection criteria and the final data sets consist of total 27 primary studies (24 SLR + 3 backward snowballing). Final set of the selected studies is provided in Appendix A, where each study is labeled as [Sn] to differentiate from the general list of references.

\subsection{Quality assessment (QA)}
\label{sec:Quality Assessment}
The assessment criteria are developed to evaluate the quality of the selected primary studies and remove the research bias. The quality assessment phase interprets the significance and completeness of each selected primary study [15]. The QA criteria checklist provided by Kitchenham and Charters [15] are analysed and designed the QA questions provided in Table \ref{tab:commands}. Each selected primary study evaluated against the quality assessment questions (QA1-QA6). Score (1) assigned if the study comprehensively addresses the quality assessment questions (see Table \ref{tab:commands}). Similarly, 0.5 points are assigned to those who have partially addressed the QA questions. Studies with no evidence of addressing the QA questions are assigned 0 points.

\begin{table*}
  \caption{Quality assessment criteria}
  \label{tab:commands}
  \begin{tabular}{c p{11cm}l c}
    \toprule
\textbf{No.} & \textbf{Assessment Questions} & \textbf{Score} \\
\midrule
    QA1	& Does the adopted research method address the research problem? & 1/0.5/0 \\
    QA2	& Does the study have clear research objectives?	& 1/0.5/0 \\
    QA3	& Does the study explicitly discuss the proposed research approach? & 1/0.5/0 \\
    QA4	& Is the study clearly reported the experimental setting? & 	1/0.5/0 \\
    QA5	& Do the study results and findings are systematically discussed? &	1/0.5/0 \\
    QA6	& Does the study present the real-world implications of the research? &	1/0.5/0 \\
     \bottomrule
  \end{tabular}
\end{table*}

\subsection{Data extraction}
\label{sec:Data Extraction}
The relevant data to address the RQs are collected by thoroughly reading the selected primary studies and extract the AI ethics principles (RQ1) and challenges (RQ2). The extracted data are recorded on excel sheets. Most of the data are collected by the second and third authors. They assess the quality of the primary studies based on the criteria discussed in Section \ref{sec:Quality Assessment}. Moreover, the first, fourth and fifth authors participated in the review meeting to finalize the QA score of each study (Appendix A).

\section{Reporting the review}
\label{sec:Reporting the review}
The data collected from the selected 27 primary studies are analyzed and discussed in the following sections.

\subsection{Temporal distribution}
\label{sec:Temporal distribution }
\begin{figure*}[!htbp]
  \centering
  \includegraphics[width=\linewidth]{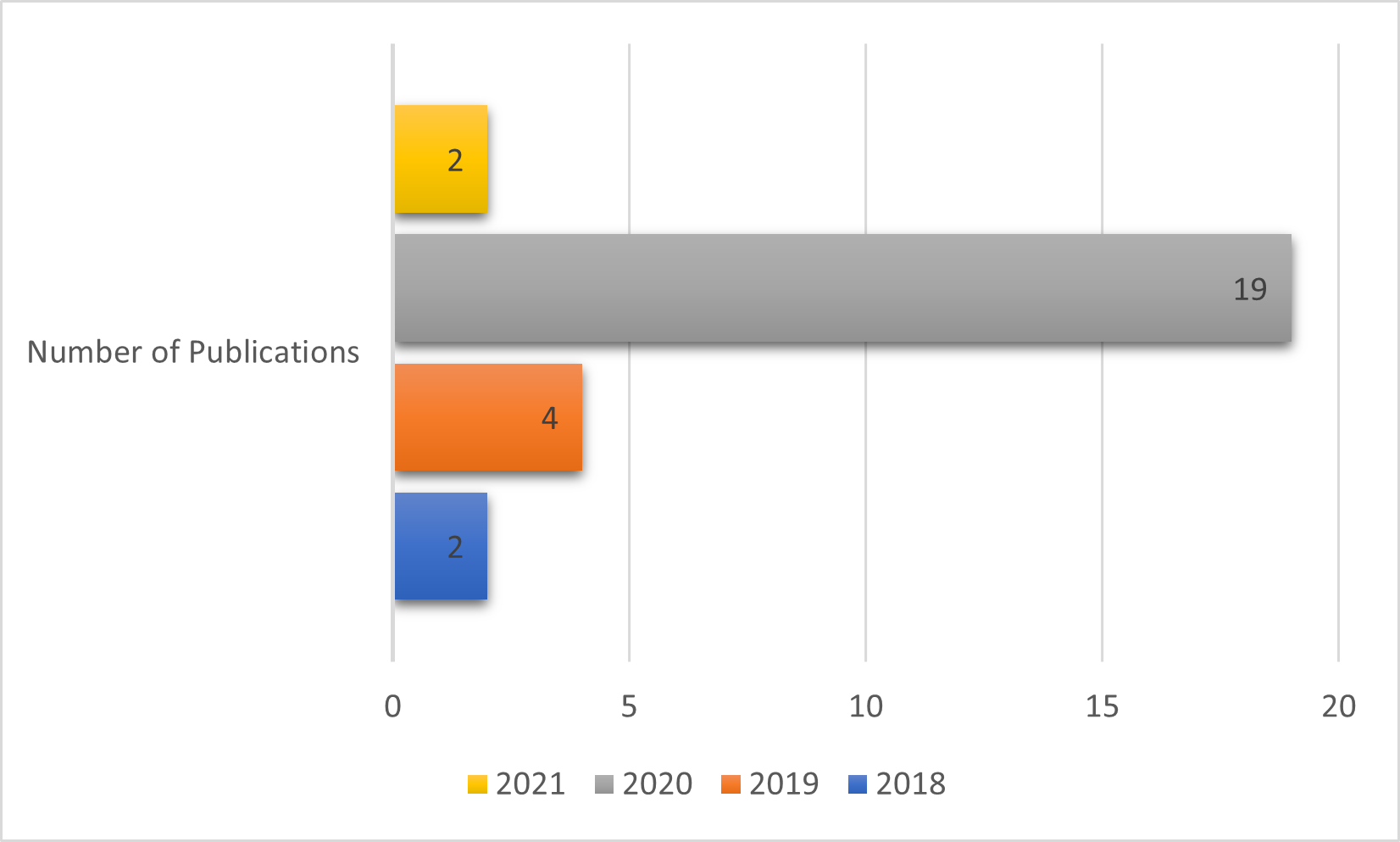}
  \caption{Temporal distribution of primary studies}
  \label{fig:TemporalDistribution}
\end{figure*}

The year wise distribution of the primary studies is shown in Fig. \ref{fig:TemporalDistribution}. Of the 27 studies, total 2, 19, 4 and 2 are respectively published in 2021 (till 5th February), 2020, 2019 and 2018. The first relevant study was found in 2018 and since then, there has been a gradual increase in the number of research publication. The SLR string was finally executed on 5th February 2021, therefore the given results only cover the first two months of 2021. The increasing number of publications reveal that AI ethics is significant, and state of the art research direction. There is still need of substantial research work to explore ethics in AI.

\subsection{Publication type}
\label{sec:Publication type}
\begin{figure*}[!htbp]
  \centering
  \includegraphics[width=\linewidth]{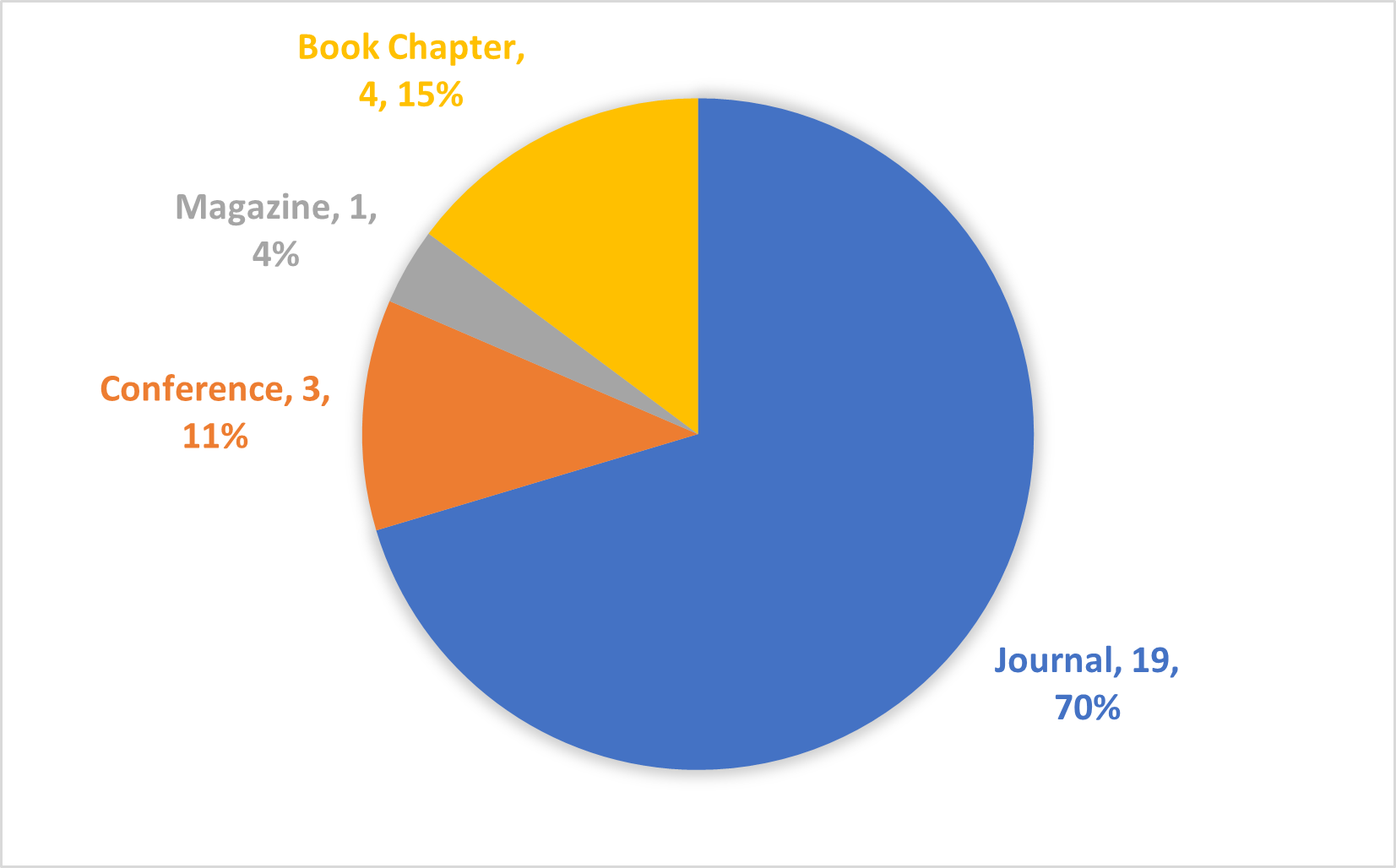}
  \caption{Publication type of primary studies}
  \label{fig:publicationstype}
\end{figure*}

The selected primary studies are classified across four major types i.e., journal, conference (including workshop), book chapter and magazine. Fig. \ref{fig:publicationstype} shows that 19 (70\%) studies are published in journals, 3 (11\%) in conferences, 4 (15\%) book chapters and 1 (4\%) is a magazine article. We noticed that journals are the most active venues to publish relevant studies.

\section{Detail results and analysis}
\label{sec:Details results and analysis}
The detail results to address RQ1 and RQ2 are discussed in the following sections.

\subsection{RQ1 (AI Ethics Principles)}
\label{sec:AI ethics principles }
\begin{figure*}[!htbp]
  \centering
  \includegraphics[width=\linewidth]{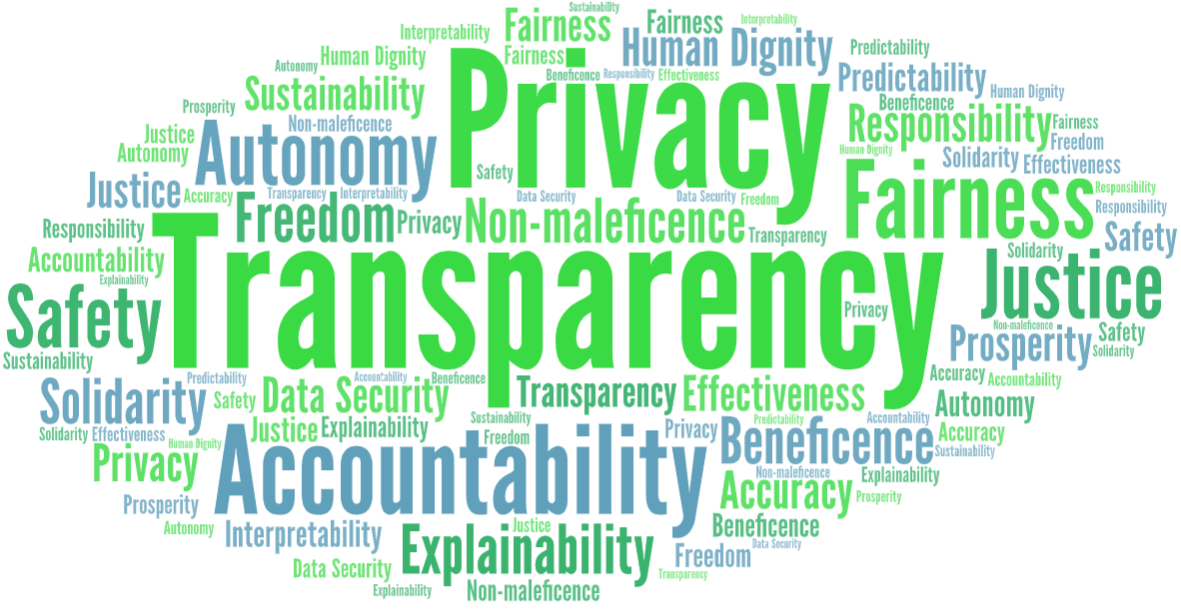}
  \caption{Word cloud of the identified AI ethics principles}
  \label{Fig:WordcloudPrinciples}
\end{figure*}

The final set of the primary studies consist of 27 articles and total 21 AI ethics principles are extracted from these articles. The identified principles along with their respective references are provided in Table \ref{tab:principles}. Moreover, a word cloud is generated to graphically represent the significance of the reported principles (See Fig. \ref{Fig:WordcloudPrinciples}). Of the 21 principles, transparency (n=17) is the most frequently mentioned principle, followed by privacy (n=16). The third and fourth most common principles are accountability (n=15) and fairness (n=14) respectively.

\begin{table*}
  \caption{Identified AI ethics principles}
  \label{tab:principles}
  \begin{tabular}{llp{6cm}cl}
    \toprule
\textbf{Principle-Id} &	\textbf{Principles} & \textbf{Reference} & \textbf{Frequency} \\
\midrule
P-01 & Transparency & [S1][S2][S3][S4][S5][S6][S7][S8][S9][S10]
[S11][S12][S13][S14][S15][S16][S17][S25] & 17 \\
P-02 & Privacy & [S1][S2][S18][S3][S5][S7][S8][S21][S22]
[S9][S11][S13][S14][S15][S16][S17] & 16 \\
P-03 & Accountability & [S1][S2][S3][S4][S5][S6][S7][S9][S10][S12]
[S13][S14][S15][S17][S25] & 15 \\
P-04 & Fairness	& [S1][S18][S4][S5][S6][S7][S8][S21] [S10][S11][S13][S14][S15][S17] & 14 \\
P-05 & Autonomy	& [S1][S2][S18][S8][S21][S22][S19][S16][S20]
[S25] & 10 \\
P-06 & Explainability & [S6][S19][S9][S11][S13][S14][S16][S20] & 8 \\
P-07  & Justice	& [S1][S18][S3][S6][S8][S19][S20] & 7 \\
P-08 & Non-maleficence & [S1][S18][S5][S6][S19][S9][S20] & 7 \\
P-09 & Human Dignity & [S1][S5][S21][S22][S10][S16] & 6 \\
P-10 & Beneficence & [S1][S18][S21][S22][S19][S20] & 6 \\
P-11 & Responsibility & [S1][S2][S5][S9][S12] & 5 \\
P-12 & Safety & [S2][S6][S7][S10][25] & 5 \\
P-13 & Data Security & [S2][S9][S16][S17] & 4 \\
P-14 & Sustainability & [S1][S6] & 2 \\
P-15 & Freedom & [S1] & 1 \\
P-16 & Solidarity & [S1] & 1 \\
P-17 & Prosperity & [S2] & 1 \\
P-18 & Effectiveness & [S2] & 1 \\
P-19 & Accuracy & [S2] & 1 \\
p-20 & Predictability & [S5] & 1 \\
P-21 & Interpretability & [S6] & 1 \\
     \bottomrule
  \end{tabular}
\end{table*}

\subsubsection{Transparency}
\label{sec:Transparency }
Transparency of operations is a major concern in AI/autonomous systems [S5]. It answers how and why a specific decision is made by the system and further triggers the other constructs including interpretability and explainability . It should not only consider for the AI system operations, but must be part of the technical process [S5] to make the decision-making actions more transparent and trustworthy. Both operational and technical transparency could be achieved by developing standards and models that measure and testify the levels of transparency. Such standards could assist the AI system development organizations to assess their level of transparency and provide best practices for further improvements. Moreover, transparency should consider for a wide range of system stakeholders; however, the level of transparency should be varied for them [S4].

\subsubsection{Privacy}
\label{sec:Privacy}
AI/autonomous system must assure user and data privacy throughout the system lifecycle. It could broadly be defined as “the right to control information about oneself” [S22]. Regulatory institutions are consistently involved in establishing legislation for data privacy and protection [S7]. However, privacy becomes more challenging in data driven AI environment, where the system subsequently processes user data including cleaning, merging, and interpretation [S7]. The data access in self-governing AI systems develop the primary concern of data privacy, which is commonly related to security and transparency [S21]. It is worth noting that AI technologies bring complex challenges associated with data privacy and integrity, which demand more relevant future research [S22].

\subsubsection{Accountability}
\label{sec:Accountability}
Accountability is the third most frequently reported principle which specifically focuses on liability issues [S5]. It refers to safeguard justice by assigning responsibility and prevent harm [S3]. The stakeholders must be accountable for the system decisions and actions to minimize the culpability problems [S4, S5]. Ensure both technical and social accountability before and after the system development, implementation and operation [S5]. Accountability is closely linked with transparency because the system must be understood before making the liability decisions [S5].

\subsubsection{Fairness}
\label{sec:Fairness}
Fairness is considered a significant principle of AI ethics. Discrimination between individuals or groups made by the decision-making systems lead to ethical fairness problems, which impact public values including dignity and justice [S11]. Avoiding unfair biases of AI systems could foster social fairness. AI and autonomous systems should not deceive people by impairing their autonomy [S4]. It could achieve by explicitly making the decision-making process more transparent and identifying the accountable entities.

\textbf{Analysis.}
Based on the SLR findings, we identified that the above principles received significant attention, which are compatible with the widely adopted accountability, responsibility and transparency (ART) framework [14] of ethics in AI. Responsibility is not a highly cited principle in the selected primary studies and the reason might be that it is considered an associated one with accountability [8]. Moreover, Vakkuri et al. [S5] developed a relational framework based on the key ART constructs with an additional fairness principle. The framework is empirically evaluated to know the opinions and perceptions of the practitioners [S5]. However, the findings of their study are only based on the five major principles and have not considered the other significant principles reported in Table \ref{tab:principles}.

\subsection{RQ2 (Challenges)}
\label{sec:RQ2 Challenges}
\begin{figure*}[!htbp]
  \centering
  \includegraphics[width=\linewidth]{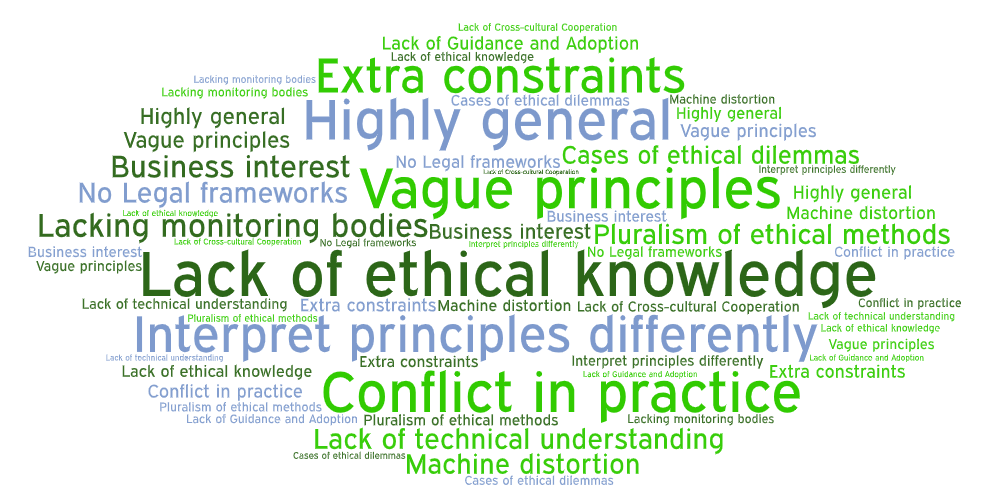}
  \caption{Word cloud of the identified AI ethics challenges}
  \label{Fig:WordCloudChallenges}
\end{figure*}

Systematic review of the 27 primary studies returns total 15 challenging factors (See Table \ref{tab:barrierstb}). The frequencies of the identified challenging factors are provided in Fig. \ref{Fig: FreqChallenges}, moreover, word cloud is generated to demonstrate the significance of the reported factors (See Fig.\ref{Fig:WordCloudChallenges}). Following are the details of the highly cited challenges:
\begin{figure*}[!htbp]
  \centering
  \includegraphics[width=\linewidth]{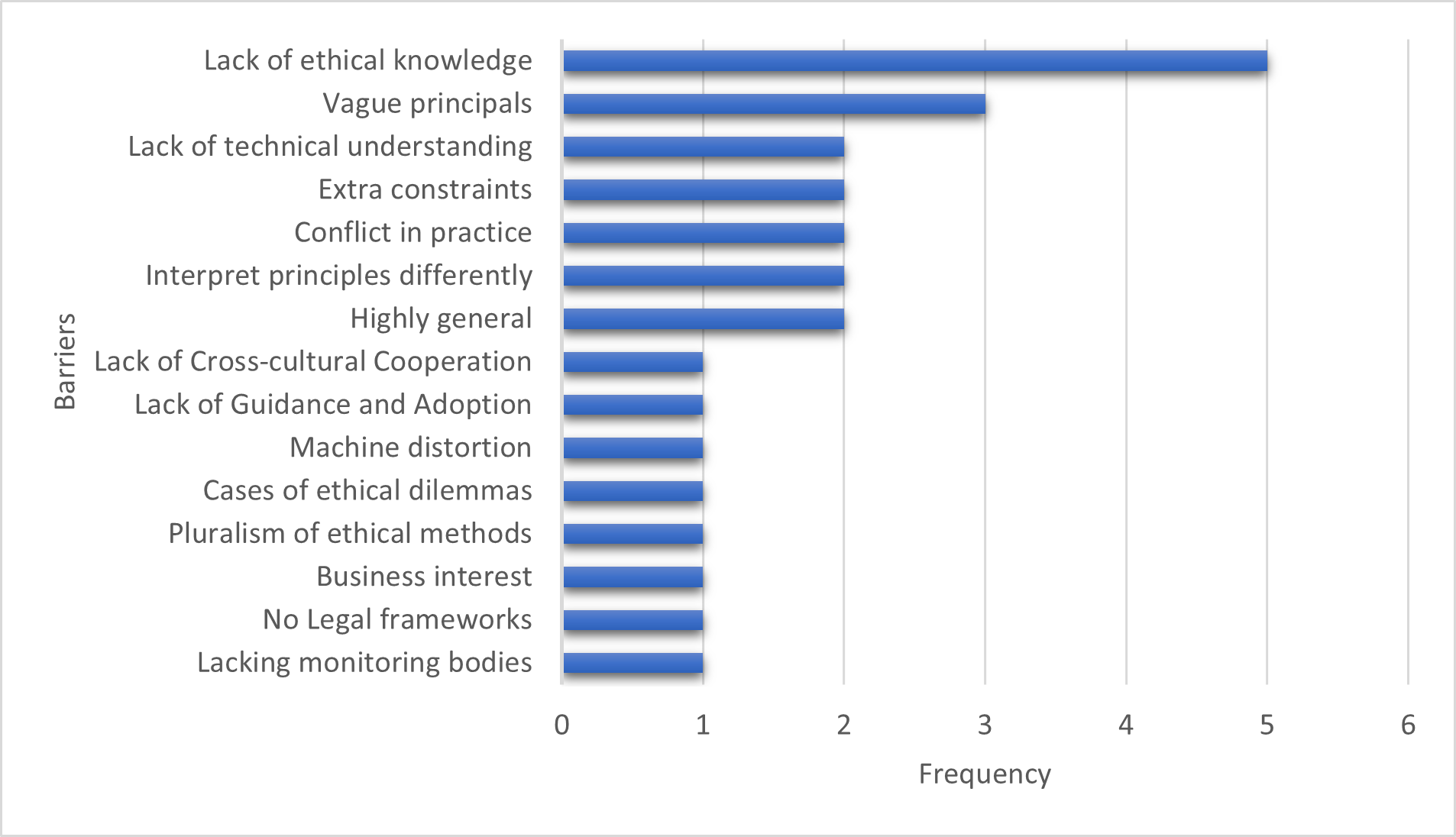}
  \caption{Frequency of identified AI ethics challenges}
  \label{Fig: FreqChallenges}
\end{figure*}

\subsubsection{Lack of ethical knowledge}
\label{sec:Lack of ethical knowledge}
Lack of ethical knowledge is one of the main reasons that AI ethics in practice is still far from being mature [S14]. AI systems development organizations believe that government institutions are not in the position of providing experts to this emerging area, while some opine that establishing ethics in AI is not possible without the political approach [S15]. Similarly, management and technical staff are not aware of the moral and ethical complexity of the AI systems. AI ethics are in their infancy, not enough ethical standards and frameworks are available that provide details guidelines to the AI industry.

\subsubsection{Vague principles}
\label{sec:Vague principles}
There are various AI ethics principles as we discussed in Section \ref{sec:AI ethics principles }. However, in practice, majority of the organizations are reluctant to adopt these principles which are highly vague in their definition [S23]. For example, it is not clear how specifically consider “fairness” and “human dignity” in AI ethics [S17]. It is very challenging to consider AI ethics in real world settings using these vaguely formulated principles [S3].

\subsubsection{Highly general}
\label{sec:Highly general}
The available principles are highly general and broad in concept to specifically consider in the AI industry [S18]. They are subjective in the term and used in various other domains than AI. Policymakers involved in drafting AI ethics principles might not have strong technical understanding of AI system development processes, which makes the principles more general and ambiguous.

\subsubsection{Conflict in practice}
\label{sec:Conflict in practices}
Organizations, committees and groups involved in developing the AI ethics guidelines and principles have opinion conflicts regarding the real world implementation of AI ethics [S13, S16]. For example, the UK house of lords suggested that robots cannot solely be in operation, but they should be guided by human beings [S10], on the other hand, in various hospitals’ robots make autonomous decisions in diagnosis and surgical endeavours. It shows interpretation and understanding conflict for AI ethics in practice.

\subsubsection{Interpret principles differently}
\label{sec:interpret principles differently}
AI ethics principles are widely considered ambiguous and general by majority of the organizations [S20]. It has been found that tech firms involved in the development of AI and autonomous systems follow ethical guidelines based on their own understandings [S27]. There are no universally agreed ethical principles that can bring all the institutions on one page.

\subsubsection{Lack of technical understanding}
\label{sec:Lack of technical understanding}
The policymakers have lack of technical knowledge, which makes AI ethics in practice a challenging effort [S10, S13]. They are not aware of the technical aspects of AI systems and the advancement in AI technologies as well their limitations. Lack of technical understanding develops the gap between system design and ethical thinking [S10]. The ethicists must have skills of grasping technical knowledge using their ethical framework [S10].

\textbf{Analysis.} The above reported challenges provide an overview of the most common and frequently cited factors that could be potential barriers for scaling ethics in AI. Lack of ethical knowledge is identified as the most common challenge of AI ethics. Major ethical mistakes are made because of no moral awareness of specific problem [S14]. Practitioners only consider software development activities as the main responsibilities; however, they have limited interest to consider ethical aspects [S5]. The ethical uncertainty in AI systems could only be diminish by acquiring ethical knowledge. Continuous awareness of ethical policies, codes and regulations assist to properly manage the ethical values in AI and autonomous systems.

We noticed that very few studies are published where the barriers of AI ethics are directly or indirectly mentioned. It is evident from the frequency distribution of the challenging factors given in Table \ref{tab:barrierstb}. This finding reveals that the AI ethics challenges aspect is very young field and requires considerable research effort from diverse disciplines to be mature. The significance of AI technologies in various sectors calls for rush research to uncover the relevant challenges that hinder the process of considering ethics in AI.

Moreover, the challenging factors having low frequency are not discussed in details because of the page limitation. However, the complete list of the identified factors is provided in Table \ref{tab:barrierstb}.

\begin{table*}
  \caption{AI ethics challenging factors}
  \label{tab:barrierstb}
  \begin{tabular}{ll p{5cm}c}
    \toprule
\textbf{Challenge-Id} &	\textbf{Challenges} & \textbf{Reference} \\
\midrule
    Ch-01 & Lack of ethical knowledge & [S14], [S15], [S18], [S21], [S23] \\
    Ch-02 & Vague principles & [S3], [S17], [S24] \\
    Ch-03 &	Highly general & [S19], [S20] \\
    Ch-04 &	Conflict in practice & [S19], [S16] \\
    Ch-05 &	Interpret principles differently & [S19], [S20] \\
    Ch-06 &	Lack of technical understanding & [S10], [S13] \\
    Ch-07 &	Extra constraints &	[S11], [S24] \\
    Ch-08 &	Lacking monitoring bodies &	[S20] \\
    Ch-09 &	No Legal frameworks	& [S20] \\
    Ch-10 &	Business interest & [S25] \\
    Ch-11 &	Pluralism of ethical methods &	[S14] \\
    Ch-12 &	Cases of ethical dilemmas & [S14] \\
    Ch-13 &	Machine distortion & [S14] \\
    Ch-14 &	Lack of Guidance and Adoption &	[S26] \\
    Ch-15 &	Lack of Cross-cultural Cooperation & [S27] \\
     \bottomrule
  \end{tabular}
\end{table*}

\begin{figure*}[!htbp]
  \centering
  \includegraphics[width=\linewidth]{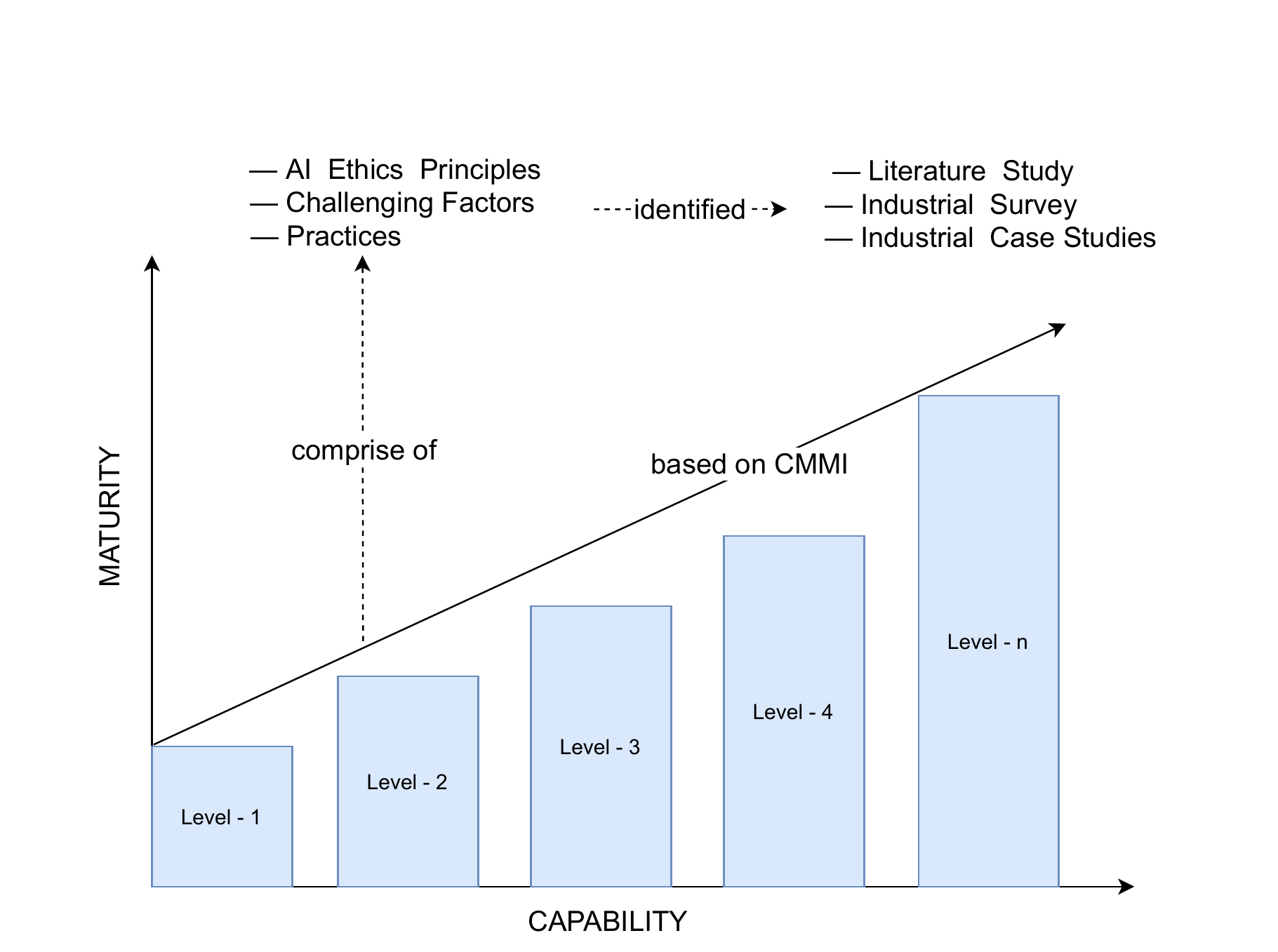}
  \caption{A proposed capability maturity model of AI ethics}
  \label{Fig: ProposedModel}
 \end{figure*}

The long-term plan of the research is to propose a maturity model that could be used to evaluate the ethical capabilities of the organizations involved in developing AI systems. The findings of this systematic review are the initial inputs for the development of the proposed model. Figure \ref{Fig: ProposedModel} shows the preliminary structure of the model and demonstrates how the findings of this review contribute in the development of the principles and challenges component. The identified principles and challenges will be classified across capability and maturity levels. Moreover, best practices will provide to tackle the identified challenges and implement the AI ethics principles. The given model is a proposed idea that will be systematically developed based on the industrial empirical studies and the concepts of the widely adopted CMMI process model [19]. Case study approach is selected to evaluate the real-world significance of the model.

\section{Threats to validity}
\label{sec:Threats to validity}
\subsection{Construct validity}
\label{sec:Construct validity}
The primary studies selection process might affect the quality of the data collected for synthesis. However, we define the formal search strategy and constantly revised it during the regular consensus meetings. Moreover, the given search string might not cover all the relevant articles and missed quality studies. Therefore, we tried to avoid this threat by conducting a pilot search using multiple strings. The final string was developed based on the results returned by the pilot strings. Finally, backward snowballing is performed to identify any additional primary studies that were missed during the SLR process.

\subsection{Internal validity}
\label{sec:Internal validity}
In SLR studies, internal validity refers to the rigorousness of the review process including the development of research questions, data sources, search strategy, study selection, string development etc. This study is conducted by following the formal SLR process guidelines proposed by Kitchenham and Charters [15]. The step-by-step flow of the SLR phases is methodically discussed in Section \ref{sec:Methodology}.

\subsection{External validity}
\label{sec:external validity}
External validity is related to the generalizability of the study findings. The results are summaries based on 27 primary studies, because of the novelty of the research topic, very few studies published in this domain. The primary studies sample size (n=27) might not be strong enough to generalize the study findings, however, we plan to extend this study in future by conducting an industrial study to evaluate the SLR findings and know the perceptions of the practitioners.

\section{Conclusions and future directions}
\label{sec:conclusions}
Ethics in AI gets significant attention in the last couple of years and there is a need of systematic literature study that discuss the principles and uncover the key challenges of AI ethics. This study is conducted to fill the given research gap by following the SLR approach. We identified total 27 relevant primary studies and the systematic review of the selected studies return 22 principles and 15 challenging factors. We noticed that most of the studies focus on four major principles i.e., transparency, privacy, accountability and fairness, which should consider by the AI system designers. Moreover, the decision-making systems should also be aware of the ethical principles to know the implications of their actions.

The challenges of ethics in AI are identified to provide an understanding of the factors that hinder the implementation of ethical principles. The most frequently reported challenging factors are lack of ethical knowledge and vague principles. The knowledge and understanding of ethics are important for both management and technical teams. It further removes the vagueness in AI principles. Lack of ethical knowledge could undermine the significance of decision-making systems.

We plan to extend this study by conducting an industrial survey to investigate the understanding of AI ethics in practice and identify the best practices to tackle the given challenging factors and manage the reported principles. Moreover, industrial case studies will be conducted in AI industry to assess the effectiveness of the proposed maturity model in practice.

\section{Acknowledgments}

\section{Appendices}
  \appendix \textbf{Appendix A: Selected primary studies}
  \begin{longtable}{l p{8cm}c lllll} 
    \toprule
\textbf{S. No.} & \textbf{Selected Primary Studies} & \textbf{Q1} & \textbf{Q2} & \textbf{Q3} & \textbf{Q4} & \textbf{Total} \\
\midrule 
S1 & Anna Jobin, Marcello Ienca, and Effy Vayena. 2019. The global landscape of AI ethics guidelines. Nature Machine Intelligence 1, no. 9 (2019), 389-399. https://doi.org/10.1038/s42256-019-0088-2 & 1 & 1 & 1 & 1 & 4 \\
S2 & Siau, Keng, and Weiyu Wang. 2020. Artificial intelligence (AI) ethics: ethics of AI and ethical AI. Journal of Database Management (JDM) 31, 2 (2020), 74-87. DOI: 10.4018/JDM.2020040105 & 1 & 1 & 1 & 1 & 4 \\
S3 & Canca Cansu. 2020. Operationalizing AI ethics principles. Communications of the ACM 63, 12 (2020), 18-21. DOI: 10.1145/3430368. & 1 & 1 & 1 & 1 & 4 \\
S4 & Hoffmann A. Lauren, Sarah T. Roberts, Christine T. Wolf, and Stacy Wood. 2018. Beyond fairness, accountability, and transparency in the ethics of algorithms: Contributions and perspectives from LIS. Proceedings of the Association for Information Science and Technology 55, 1 (2018), 694-696. https://doi.org/10.1002/pra2.2018.14505501084 & 1 & 1 & 0 & 0.5 & 2.5 \\
S5 & Ville Vakkuri, Kai-Kristian Kemell, Joni Kultanen and Pekka Abrahamsson. 2020. The Current State of Industrial Practice in Artificial Intelligence Ethics. IEEE Software 37, 4 (2020), 50-57. DOI: 10.1109/MS.2020.2985621 & 1 & 1 & 1 & 1 & 4 \\
S6 & Charles D. Raab. 2020. Information privacy, impact assessment, and the place of ethics. Computer Law \& Security Review 37 (2020), 105404. https://doi.org/10.1016/j.clsr.2020.105404 & 1 & 1 & 1 & 1 & 4 \\
S7 & Wenjun Wu, Tiejun Huang and Ke Gong. 2020. "Ethical principles and governance technology development of AI in China. Engineering 6, 3 (2020), 302-309. https://doi.org/10.1016/j.eng.2019.12.015 & 1 & 1 & 1 & 1 & 4 \\
S8 & Sara Gerke, Timo Minssen, and Glenn Cohen. 2020. Ethical and legal challenges of artificial intelligence-driven healthcare. In Artificial intelligence in healthcare, (2020), 295-336. https://doi.org/10.1016/B978-0-12-818438-7.00012-5 & 1 & 0.5 &	1 & 0.5 & 3 \\
S9 & Richard Benjamins. 2021. A choices framework for the responsible use of AI. AI and Ethics 1, 1 (2021), 49-53. https://doi.org/10.1007/s43681-020-00012-5 & 1 & 1 & 1 & 1 & 4 \\
S10 & Thilo Hagendorff. 2020. The ethics of AI ethics: An evaluation of guidelines. Minds and Machines 30, no. 1 (2020), 99-120. https://doi.org/10.1007/s11023-020-09517-8 & 1 & 1 & 1 & 1 & 4 & \\
S11 & Nagadivya Balasubramaniam, Marjo Kauppinen, Sari Kujala and Kari Hiekkanen. 2020. Ethical Guidelines for Solving Ethical Issues and Developing AI Systems. In International Conference on Product-Focused Software Process Improvement, Lecture Notes in Computer Science, vol 12562, Springer, Cham, 331-346. https://doi.org/10.1007/978-3-030-64148-1\_21 & 1 &	1 &	1 &	1 &	4 \\
S12 &	Ville Vakkuri and Kai-Kristian Kemell. 2019. Implementing AI Ethics in Practice: An Empirical Evaluation of the RESOLVEDD Strategy. In International Conference on Software Business. Lecture Notes in Business Information Processing, vol 370. Springer, Cham. https://doi.org/10.1007/978-3-030-33742-1\_21 pp. 260-275 & 1 &	1 &	1 & 1 & 4 \\
S13 & Ray Eitel-Porter. 2021. Beyond the promise: implementing ethical AI. AI and Ethics 1, 1 (2021), 73-80. https://doi.org/10.1007/s43681-020-00011-6 & 1 & 1 & 1 & 1 & 4 \\ 
S14 & Tomas Hauer. 2020. Machine Ethics, Allostery and Philosophical Anti-Dualism: Will AI Ever Make Ethically Autonomous Decisions?." Society 57, 4 (2020), 425-433. https://doi.org/10.1007/s12115-020-00506-2 &  1 & 1 & 1 & 1 & 4 \\
S15 & Josef Baker-Brunnbauer. 2020. Management perspective of ethics in artificial intelligence." AI and Ethics (2020), 1-9. https://doi.org/10.1007/s43681-020-00022-3 & 1 & 1 & 1 & 1 & 4 \\
S16 & Banu Buruk, Perihan Elif Ekmekci and Berna Arda. 2020. A critical perspective on guidelines for responsible and trustworthy artificial intelligence. Medicine, Health Care and Philosophy 23, 3 (2020), 387-399. https://doi.org/10.1007/s11019-020-09948-1 & 1 & 	1 &	1 &	1 &	4 \\
S17 & Mittelstadt, Brent. 2019. Principles alone cannot guarantee ethical AI. Nature Machine Intelligence 1, 11 (2019), 501-507. https://doi.org/10.1038/s42256-019-0114-4 & 1 &	1 &	1 &	1 &	4 \\
S18 & Jess Whittlestone, Rune Nyrup, Anna Alexandrova, and Stephen Cave. 2019. The Role and Limits of Principles in AI Ethics: Towards a Focus on Tensions. In Proceedings of the 2019 AAAI/ACM Conference on AI, Ethics, and Society (AIES '19). Association for Computing Machinery, New York, NY, USA, 195–200. DOI:https://doi.org/10.1145/3306618.3314289 & 1	& 1 & 0 & 1 & 3 \\
S19 &	Luciano Floridi, Josh Cowls, Monica Beltrametti, Raja Chatila, Patrice Chazerand, Virginia Dignum, Christoph Luetge, Robert Madelin, Ugo Pagallo, Francesca Rossi, Burkhard Schafer, Peggy Valcke and Effy Vayena. 2018.  AI4People—an ethical framework for a good AI society: opportunities, risks, principles, and recommendations. Minds and Machines 28, 4 (2018), 689-707. https://doi.org/10.1007/s11023-018-9482-5 & 1 &	1 &	1 &	1 &	4 \\
S20 & Jessica Morley, Luciano Floridi, Libby Kinsey, and Anat Elhalal. 2020. From what to how: an initial review of publicly available AI ethics tools, methods and research to translate principles into practices. Science and engineering ethics 26, 4 (2020), 2141-2168. https://doi.org/10.1007/s11948-019-00165-5 & 1 & 1 &	1 &	1 &	4 \\
S21 & Ángel G. de Ágreda. 2020. Ethics of autonomous weapons systems and its applicability to any AI systems. Telecommunications Policy 44, 6 (2020), 101953. https://doi.org/10.1016/j.telpol.2020.101953 & 1 & 1 & 1 & 1 & 4 \\
S22 &	Shuili Du, and Chunyan Xie. 2020. Paradoxes of artificial intelligence in consumer markets: Ethical challenges and opportunities. Journal of Business Research (2020). https://doi.org/10.1016/j.jbusres.2020.08.024. & 1	& 1 & 1 & 1 &	4 \\
S23 & Mark Coeckelbergh. 2020. 11 CHALLENGES FOR POLICYMAKERS, AI Ethics, MIT Press, 2020, 167-181. DOI: 10.7551/mitpress/12549.001.0001	& 0.5 &	0.5 & 1 & 1	& 3 \\
S24 & E. L. Sidorenko, Z. I. Khisamova and U. E. Monastyrsky. 2021. The Main Ethical Risks of Using Artificial Intelligence in Business. In Current Achievements, Challenges and Digital Chances of Knowledge Based Economy, Springer, Cham, 423-429. https://doi.org/10.1007/978-3-030-47458-4\_51 & 0.5 & 1 & 1 & 0.5 & 3 \\
S25 & Chris Rees. 2020 .The Ethics of Artificial Intelligence. In Unimagined Futures–ICT Opportunities and Challenges, Springer, Cham, 55-69. https://doi.org/10.1007/978-3-030-64246-4\_5	& 1 & 1 & 1 & 1 & 4 \\
S26 & Jan Jöhnk, Malte Weißert and Katrin Wyrtki. 2021. Ready or Not, AI Comes—An Interview Study of Organizational AI Readiness Factors. Business \& Information Systems Engineering 63, 1 (2021), 5-20. https://doi.org/10.1007/s12599-020-00676-7 &  1 &	1 &	1 &	1 & 4 \\
S27 & Seán S. ÓhÉigeartaigh, Jess Whittlestone, Yang Liu, Yi Zeng, and Zhe Liu. 2020. Overcoming barriers to cross-cultural cooperation in AI ethics and governance. Philosophy \& Technology 33, 4 (2020), 571-593. https://doi.org/10.1007/s13347-020-00402-x & 1 & 1 & 1 & 1 & 4 \\

\bottomrule

  \end{longtable}

\begin{thebibliography}{00}
\bibitem{1}  Christina Pazzanese. 2020. Ethical concerns mount as AI takes bigger decision-making role in more industries. Retrieved January 15, 2021 from  https://news.harvard.edu/gazette/story/2020/10/ethical-concerns-mount-as-ai-takes-bigger-decision-making-role/
\bibitem{2} Vincent C. Müller. 2020. Ethics of Artificial Intelligence and Robotics. The Stanford Encyclopedia of Philosophy. Retrieved January 15, 2021 from https://plato.stanford.edu/archives/win2020/entries/ethics-ai/
\bibitem{3} Ville Vakkuri, Kai-Kristian Kemell and Pekka Abrahamsson. 2019. Implementing Ethics in AI: Initial Results of an Industrial Multiple Case Study. In International Conference on Product-Focused Software Process Improvement, Lecture Notes in Computer Science, vol 11915. Springer, Cham, 331-338. https://doi.org/10.1007/978-3-030-35333-9\_24
\bibitem{4} Jaana Leikas, Raija Koivisto, and Nadezhda Gotcheva. 2019. Ethical framework for designing autonomous intelligent systems. Journal of Open Innovation: Technology, Market, and Complexity 5, 1 (2019), 18. https://doi.org/10.3390/joitmc5010018
\bibitem{5} Ville Vakkuri, Kai-Kristian Kemell and Pekka Abrahamsson. 2019. AI ethics in industry: a research framework. arXiv preprint arXiv:1910.12695.
\bibitem{6} IEEE. (2019). Ethically aligned design: A vision for prioritizing human well-being with autonomous and intelligent systems, first edition. Retrieved January 17, 2021 from  https://tinyurl.com/yah4jzb6
\bibitem{7} Anna Jobin, Marcello Ienca, and Effy Vayena. 2019. The global landscape of AI ethics guidelines. Nature Machine Intelligence 1, no. 9 (2019), 389-399. https://doi.org/10.1038/s42256-019-0088-2
\bibitem{8} Pekka Ala-Pietilä, Wilhelm Bauer, Urs Bergmann, Mária Bieliková, Cecilia Bonefeld-Dahl, Yann Bonnet, Loubna Bouarfa et al. (2018). The European Commission’s high-level expert group on artificial intelligence: Ethics guidelines for trustworthy AI. Working Document for stakeholders’ consultation. Retrieved January 17, 2021 from https://ec.europa.eu/digital-single-market/en/news/ethics-guidelines-trustworthy-ai
\bibitem{9} Alan Bundy. 2016. Preparing for the future of artificial intelligence. Executive Office of the President National Science and Technology Council Committee on Technology Washington, D.C, USA. Retrieved January 23, 2021 from https://cra.org/ccc/wp-content/uploads/sites/2/2016/11/NSTC\_preparing\_for\_the\_future\_of\_ai.pdf 
\bibitem{10} Beijing Academy of Artificial Intelligence. 2019. Beijing AI principles. Retrieved January 23, 2021 from https://www.baai.ac.cn/news/beijing-ai-principles-en.html
\bibitem{11} ISO/IEC. ISO/IEC JTC 1/SC 42 Artificial intelligence, Retrieved 25th January 2021, https://www.iso.org/committee/6794475.html
\bibitem{12} Ville Vakkuri, Kai-Kristian Kemell, Marianna Jantunen, and Pekka Abrahamsson. 2020. “This is Just a Prototype”: How Ethics Are Ignored in Software Startup-Like Environments. In International Conference on Agile Software Development, Lecture Notes in Business Information Processing, vol 383. Springer, Cham. 195-210. https://doi.org/10.1007/978-3-030-49392-9\_13
\bibitem{13} Andrew McNamara, Justin Smith, and Emerson Murphy-Hill. 2018. Does ACM’s code of ethics change ethical decision making in software development? In Proceedings of the 26th ACM Joint Meeting on European Software Engineering Conference and Symposium on the Foundations of Software Engineering (ESEC/FSE 2018). Association for Computing Machinery, New York, NY, USA, 729–733. DOI:https://doi.org/10.1145/3236024.3264833
\bibitem{14} Virginia Dignum. 2017. Responsible autonomy. Preprint arXiv:1706.02513 
\bibitem{15} Barbara Kitchenham and Stuart Charters. 2007. Guidelines for performing systematic literature reviews in software engineering. Technical report, Ver. 2.3 EBSE Technical Report. School of Computer Science and Mathematics, Keele University, UK.
\bibitem{16} Lianipng Chen, Muhammad Ali Babar, and He Zhang. 2010. Towards an evidence-based understanding of electronic data sources. In Proceedings of the 14th international conference on Evaluation and Assessment in Software Engineering (EASE'10). BCS Learning \& Development Ltd., Swindon, GBR, 135–138.
\bibitem{17} Claes Wohlin. 2014. Guidelines for snowballing in systematic literature studies and a replication in software engineering. In Proceedings of the 18th International Conference on Evaluation and Assessment in Software Engineering (EASE '14). Association for Computing Machinery, New York, USA, 1–10. DOI:https://doi.org/10.1145/2601248.2601268
\bibitem{18} Tingting Bi, Peng Liang, Antony Tang, and Chen Yang. 2018. A systematic mapping study on text analysis techniques in software architecture. Journal of Systems and Software 144, (2018), 533-558.https://doi.org/10.1016/j.jss.2018.07.055 
\bibitem{19} CMMI Product Team. 2002. Capability maturity model® integration (CMMI SM), version 1.1. CMMI for systems engineering, software engineering, integrated product and process development, and supplier sourcing (CMMI-SE/SW/IPPD/SS, V1. 1) 2 (2002).


\end{thebibliography}
\end{document}